\documentclass[twocolumn,twoside,slac_two]{revtex4}
\usepackage{graphicx}
\usepackage{fancyhdr}
\begin{document}

\title{New limit of pion form factor at very large $Q^2$}

%

\author{B.A.Li}
\affiliation{Department of Physics and Astronomy, University of Kentucky, Lexington, KY 40506, USA}

\begin{abstract}
A new limit of pion form factor at very large $Q^2$ is obtained by using a pion wave function determined from an
effective chiral field theory of mesons.
It shows that
when $Q^2>>(1.8GeV)^2$ the pion form factor reaches the asymptotic limit ${\alpha_s(Q^2)\over Q^2}$.
\end{abstract}

\maketitle

\thispagestyle{fancy}




Pion form factor is a very important quantity in hadron physics.

In 1973 the quark counting rule~\cite{br1} predicts
\begin{equation}
F_\pi(Q^2)\sim \frac{1}{Q^2}.
\end{equation}
Perturbative QCD~\cite{br2} predicts that dominance of one gluon exchange at large $Q^2$
\begin{equation}
F_\pi(Q^2)|_{Q^2\rightarrow\infty}=4\pi\alpha_s(Q^2)f^2_\pi/Q^2.
\end{equation}
The issue in the study of $F_\pi(Q^2)$ is that the pion wave function or the distribution
amplitude is a quantity of nonperturbative QCD.
Other different pion distribution amplitudes
are discussed in the light-cone formalism too. For example,
\[\{x(1-x)\}^{{1\over2}}\]
this new function increases the
the value of $F_\pi(Q^2)$ at $Q^2\rightarrow\infty$ by ${16\over9}$ relative to the prediction obtained
by x(1-x).
Different distribution amplitude leads to different
asymptotic value of $F_\pi(Q^2)$ at $Q^2\rightarrow\infty$.
The determination of pion wave function or distribution amplitude is still a open question.
In this talk another BS wave function of pion is introduced to calculate $F_\pi(Q^2)$ at large
$Q^2$.

According to Mandelstam's representation, the current matrix element of pion is written as
\begin{eqnarray}
<\pi^+|j_\mu(0)|\pi^+>=\int d^4 k_2\int d^4 k_1 Tr\{\phi_\pi(k_1,p_f)a\nonumber
\end{eqnarray}
\begin{eqnarray}
T_H(k_1,k_2,p_f,p_i)_\mu
\phi_\pi(k_2,p_i)\}
=F_\pi(Q^2)P_\mu.
\end{eqnarray}
The same
pion wave function should determine $F_\pi(Q^2)$ at both low $Q^2$ and high
$Q^2$.

Current algebra is successful
in studying hadron physics at lower energies, in which chiral symmetry plays essential role.
Based on QCD and current algebra an effective chiral Lagrangian
of pseudoscalar, vector, and axial-vector mesons is constructed~\cite{li1} as
\begin{eqnarray}
{\cal L}=\bar{\psi}(x)(i\gamma\cdot\partial+\gamma\cdot v
+\gamma\cdot a\gamma_{5}
-mu(x))\psi(x)\nonumber
\end{eqnarray}
\begin{eqnarray}
+{1\over 2}m^{2}_{0}(\rho^{\mu}_{i}\rho_{\mu i}+
\omega^{\mu}\omega_{\mu}+a^{\mu}_{i}a_{\mu i}+f^{\mu}f_{\mu}),
\end{eqnarray}
where \(a_{\mu}=\tau_{i}a^{i}_{\mu}+f_{\mu}\), \(v_{\mu}=\tau_{i}
\rho^{i}_{\mu}+\omega_{\mu}\),
\(u=exp\{i\gamma_{5}(\tau_{i}\pi_{i}+
\eta)\}\), 
m is the constituent quark mass and it originates in quark condensation. Therefore,
this theory has dynamical chiral symmetry breaking.
Integrating out the quark fields, the Lagrangian
of mesons is obtained.
The tree diagrams of mesons are at leading order in $N_C$ expansion and loop diagrams of
mesons are at higher order. In the limit, $m_q\rightarrow0$, explicit chiral symmetry
is recovered.
It is known that chiral symmetry, quark condensation and $N_C$ expansion are from
nonperturbative QCD.
Meson physics at lower energies has been extensively studied by this Lagrangian. Theory agrees with
data very well.

The pion decay constant is defined
\[f^2_\pi=F^2(1-{2c\over g}),\]
\[{F^2\over16}=\frac{N_C}{(4\pi)^2}\int d^4k\frac{m^2}{(k^2+m^2)^2},\]
c is determined to be \(c=\frac{f^2_\pi}{2gm^2_\rho}\),
\[g^2={2\over3}\frac{N_C}{(4\pi)^4}\int d^4k\frac{1}{(k^2+m^2)^2}={1\over6}{F^2\over m^2}.\]
$f_\pi$ and g are the two parameters of this theory. \(g=0.39\)
is determined from the decay rate of $\rho\rightarrow ee^+$.
The cut-off $\Lambda$ of the integrals is determined to be $\Lambda=1.8GeV$.

The pion form factor is derived  up to the fourth order in covariant derivatives~\cite{ti2}
\[|F_\pi (q^2)|^2 = f^2_{\rho \pi \pi }(q^2)
\frac{m_\rho ^4+q^2\Gamma _\rho
^2(q^2)}{(q^2-m_\rho ^2)^2+q^2\Gamma _\rho ^2(q^2)},\]
\[f_{\rho \pi \pi }(q^2)= 1+\frac{q^2}{2\pi ^2f_\pi
^2}[(1-\frac{2c}{g})^2-4\pi ^2c^2],\]
\[\Gamma_\rho(q^2)=\frac{f^2_{\rho\pi\pi}(q^2)}{12\pi g^2}(1-{4m^2_{\pi}\over q^2})^{{3\over2}}.\]
In these equations there is no new adjustable parameter.
At $q^2=m^2_\rho$ \(\Gamma_\rho=150MeV\) which is consistent with data.
$F_\pi (q^2)$ consists of two parts: the $\rho$ pole and an intrinsic
form factor $f_{\rho \pi \pi }(q^2)$ which is obtained from quark loop.
The intrinsic form factor is a new result of this theory.
The $\rho$-pole form factor of pion has
shortcomings: in space-like region it decreases too slow and in time-like region it decreases
too fast. The intrinsic form factor redeems these two problems. The comparison between the theory and the data
is shown

\begin{figure}[h]
\centering
\includegraphics[width=80mm]{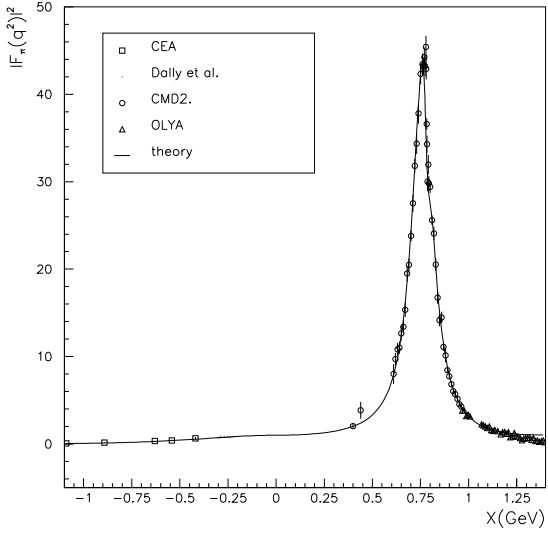}
\caption{Pion form factor} 
\end{figure}
The radius of charged pion is obtained
\begin{eqnarray}
<r^2>_\pi={6\over m^2_\rho}+{3\over \pi^2 f^2_\pi}\{(1-{2c\over g})^2-4\pi^2 c^2\}\nonumber
\end{eqnarray}
\begin{eqnarray}
=0.452 fm^2.
\end{eqnarray}
The experimental data is $(0.439\pm0.03)fm^2$.
The contribution of the $\rho$ pole to $<r^2>_\pi$ is $0.395fm^2$.

In the leading order in $N_C$ expansion(ignore the loop diagrams of mesons) a pion is made of constituent
quark pair with $\rho$ meson cloud. The intrinsic form factor is obtained from the quark pair component
and the $\rho-pole$ of $F_\pi(q^2)$ is from the $\rho$ meson cloud.

The BS wave function of pion can be defined in this theory.
The pion fields have two sources: \(u=exp\{i\gamma_{5}\pi\}=
1+i\gamma_{5}\pi+...\) and the shifting $a_\mu\rightarrow a_\mu(physical)-{c\over g}\partial_\mu\pi$,
which is caused by the mixing term $a^i_\mu\partial_\mu\pi^i$ obtained from quark loop diagram.
Combining these two sources together, the vertex related to pion is obtained as
\begin{eqnarray}
{\cal L}_\pi=-{2im\over f_\pi}\bar{\psi}\tau_i\gamma_5(1+i{c\over g}{\gamma\cdot\partial\over m})
\psi\pi_i.
\end{eqnarray}
The wave function of pion is derived as
\begin{eqnarray}
\phi(z,p)=<0|\{\psi({z\over2})\bar{\psi}(-{z\over2})\}|\pi(p)>\nonumber
\end{eqnarray}
\begin{eqnarray}
=\frac{2\sqrt{2}m}{f_\pi}\frac{1}{(2\pi)^4}
\int d^4k\frac{e^{-ikz}}{(k^2-m^2)((k-p)^2-m^2)}\nonumber
\end{eqnarray}
\begin{eqnarray}
(\gamma\cdot k+m)\gamma_5\nonumber
\end{eqnarray}
\begin{eqnarray}
(1+{c\over g}{\gamma\cdot p\over m})
(\gamma\cdot k-\gamma\cdot p+m).
\end{eqnarray}
$f_\pi$ can be derived by the pion wave function too
\begin{eqnarray}
Tr\phi(0,p)\gamma_\mu\gamma_5={i\over \sqrt{2}}f_\pi p_\mu.
\end{eqnarray}

The pion mass has been derived as
\begin{equation}
m^2_\pi=-{4\over f^2_\pi}<\bar{\psi}\psi>(m_u+m_d).
\end{equation}
In the chiral limit, $m_q\rightarrow 0$, $m^2_\pi\rightarrow 0$. Goldstone theorem is satisfied.
The three form factors of $\pi^-\rightarrow e\gamma\nu$, $\pi-\pi$ scattering,
$\pi^0\rightarrow\gamma\gamma$, and many other processes in which pion is involved are
successfully studied by this theory.

The pion wave function determined by this
theory is successful in studying pion physics at lower energies. Now this wave function is used
to study pion form factor at large $Q^2$.
At large $Q^2$ perturbative QCD is working.
Using the pion wave function and $T_H$ with one gluon exchange
the matrix element of current of pion is written as
\begin{eqnarray}
<\pi^+|j_\mu|\pi^+>={2m^2\over f^2_\pi}Tr\lambda^a \lambda^a g^2_s\int d^4 k_1 d^4 k_2\nonumber
\end{eqnarray}
\begin{eqnarray}
\{\frac{1}{(k_1-k_2
+p_i-p_f)^2}\frac{1}{(k_1+p_i-p_f)^2}Tr\gamma_\nu\phi_\pi(k_1,p_f)\nonumber
\end{eqnarray}
\begin{eqnarray}
\gamma_\mu \gamma\cdot(k_1+p_i-p_2)
\gamma_\nu\phi_\pi(k_2,p_i)\nonumber
\end{eqnarray}
\begin{eqnarray}
+\frac{1}{(k_1-k_2+p_i-p_f)^2}\frac{1}{(k_2+p_f-p_i)^2}Tr\gamma_\nu\phi_\pi(k_1,p_f)\nonumber
\end{eqnarray}
\begin{eqnarray}
\gamma_\nu\gamma\cdot
(k_2+p_f-p_i)\gamma_\mu\phi_\pi(k_2,p_i)\},
\end{eqnarray}
where $\phi_\pi(k,p)$ is the pion wave function in momentum space, k is the internal momentum and p is
the momentum of the pion,
\begin{eqnarray}
\phi_\pi(k_1,p_f)=\frac{1}{(2\pi)^4}\frac{1}{(k^2_1-m^2)[(k_1-p_f)^2-m^2]}\nonumber
\end{eqnarray}
\begin{eqnarray}
[\gamma\cdot(k_1-p_f)+m]
\gamma_5(1-{c\over g}{\gamma\cdot p_f\over m})(\gamma\cdot k_1+m),\nonumber
\end{eqnarray}
\begin{eqnarray}
\phi_\pi(k_2,p_i)=\frac{1}{(2\pi)^4}\frac{1}{(k^2_1-m^2)[(k_2-p_i)^2-m^2]}\nonumber
\end{eqnarray}
\begin{eqnarray}
(\gamma\cdot k_2+m)\gamma_5
(1+{c\over g}{\gamma\cdot p_i\over m})[\gamma\cdot(k_2-p_i)+m].
\end{eqnarray}

For $Q^2>>\Lambda^2((1.8GeV)^2)$
\begin{eqnarray}
<\pi^+|j_\mu|\pi^+>={2m^2\over f^2_\pi}Tr\lambda^a \lambda^a g^2_s
{1\over Q^4}\int d^4 k_1 d^4 k_2\nonumber
\end{eqnarray}
\begin{eqnarray}
\{
Tr\gamma_\nu\phi_\pi(k_1,p_f)\gamma_\mu \gamma\cdot(k_1+p_i-p_2)
\gamma_\nu\phi_\pi(k_2,p_i)\nonumber
\end{eqnarray}
\begin{eqnarray}
+Tr\gamma_\nu\phi_\pi(k_1,p_f)\gamma_\nu\gamma\cdot
(k_2+p_f-p_i)\gamma_\mu\phi_\pi(k_2,p_i)\}
\end{eqnarray}
and the pion form factor at $Q^2>>(1.8GeV)^2$ is obtained
\begin{eqnarray}
F_\pi(Q^2)=4\pi\alpha_s(Q^2)f^2_\pi{1\over Q^2}{1\over18}(1-{2c\over g})^{-2}
\end{eqnarray}
\begin{eqnarray}
\{{2c^2\over g^2}
+(1-{c\over g})(1-{4c\over g})-{1\over 4\pi^2 g^2}(1-{c\over g})(1-{2c\over g})\}.
\end{eqnarray}
The numerical result is
\begin{equation}
F_\pi(Q^2)=2.65\times10^{-2}4\pi\alpha_s(Q^2)f^2_\pi{1\over Q^2}.
\end{equation}
It is interesting to mention that at high $Q^2$ the $\rho$-pole with one gluon exchange behaves
like ${1\over Q^4}$. Therefore, at high $Q^2$ the contribution of $\rho$-pole can be ignored.

The numerical value
of the coefficient of the new limit of the pion form factor is much smaller than previous one.
It can be tested by
experimental measurements of the $F_\pi(Q^2)$ at very large $Q^2$.








\begin{thebibliography}{9}   

\bibitem{br1} S.J.Brodsky and G.R.Farra, Phys.Rev.Lett. {\bf 31},1153(1973);
           V.A.Mateev, R.M.Muradyan and A.N.Tavhelidze, Lett. Nuovo Cimento, {\bf 7},719(1973).
\bibitem{br2} (a)G.R.Farrar and D.R.Jackson, Phys.Rev.Lett., {\bf 43}, 246(1979);
           (b)G.P.Lepage and S.J.Brodsky, Phys. Lett., {\bf B87},359(1979);
           (c)A.V.Efremov and A.V.Radyushkin, Phys. Lett.{\bf B94},245(1980)and Theoretical and
              Mathmetical Phys., {\bf 42},147(1980).
           {\bf B381},129(1992).
\bibitem{li1} B.A.Li, Phys.Rev.{\bf D52},5165-5183(1995); {\bf D52},5184-5193(1995); see review article
             B.A.Li, Inern. J of Modern Physics, {\bf A21}, 9509(2006).
\bibitem{ti2} J.Gao and B.A.Li, Phys.Rev.{\bf D61}, 113006(2000);B.A.Li and J.X.Wang, Phys.Lett.
           {\bf B543},48 (2002).

\end{thebibliography}
\end{document}